\documentclass[10pt,letterpaper,twocolumn]{article} 

\usepackage{ol2}
\usepackage[draft]{hyperref}
\usepackage{amsmath}

\usepackage{units}
\usepackage[utf8]{inputenc}
\usepackage{amssymb}

\newcommand{\w}{_{\mathrm{w}}}
\renewcommand{\c}{_{\mathrm{c}}}
\newcommand{\cw}{_{\mathrm{cw}}}

\newcommand{\ww}{_{\mathrm{ww}}}

\newcommand{\ac}{_{\mathrm{ac}}}

\newcommand{\um}{\mu m}

\newcommand{\mytilde}{\raise.17ex\hbox{$\scriptstyle\sim$}}

\begin{document}

\twocolumn[

\title{Array of Fabry-Pérot waveguide resonators with tunable coupling}

\author{Guillaume Lepert,$^{1,*}$ E.A. Hinds,$^{1,*}$ Helen L. Rogers,$^2$ James C. Gates$^2$ and Peter G.R. Smith$^{2}$}

\address{
$^1$Centre for Cold Matter, Blackett Laboratory, Imperial College, London SW7 2BW, UK\\
$^2$Optoelectronics Research Centre, University of Southampton, Southampton, SO17 1BJ, UK\\
$^*$Corresponding authors: guillaume.lepert07@imperial.ac.uk, ed.hinds@imperial.ac.uk
}

\begin{abstract}
	We demonstrate the elements of a coupled-resonator optical waveguide in a side-coupled Fabry-Pérot configuration, and show that the coupling rate between adjacent waveguides can be widely tuned through the thermo-optic effect. The device is linearly scalable and can be combined with other integrated devices, with applications as an optical delay line or as a key element in a cavity-QED based quantum simulator.
\end{abstract}


 ]


Coupled-resonator optical waveguides (CROWs) are of great topical interest in the telecom industry as tunable delay lines. Today most of these devices consist of a chain of ring resonators based on silicon and other high-index waveguides \cite{Morichetti2012}, where the delay can be tuned by switching the rings out of resonance \cite{Melloni2008}. The couplings are not adjustable, being determined by the ring spacings, which therefore demand very careful engineering to achieve the desired spectral response of the filter. In this letter we propose and demonstrate a novel mechanism to tune the couplings in real time using  a different type of CROW, based on side-coupled waveguide Fabry-Pérot cavities.

\begin{figure}[b]
	\includegraphics[width=\columnwidth]{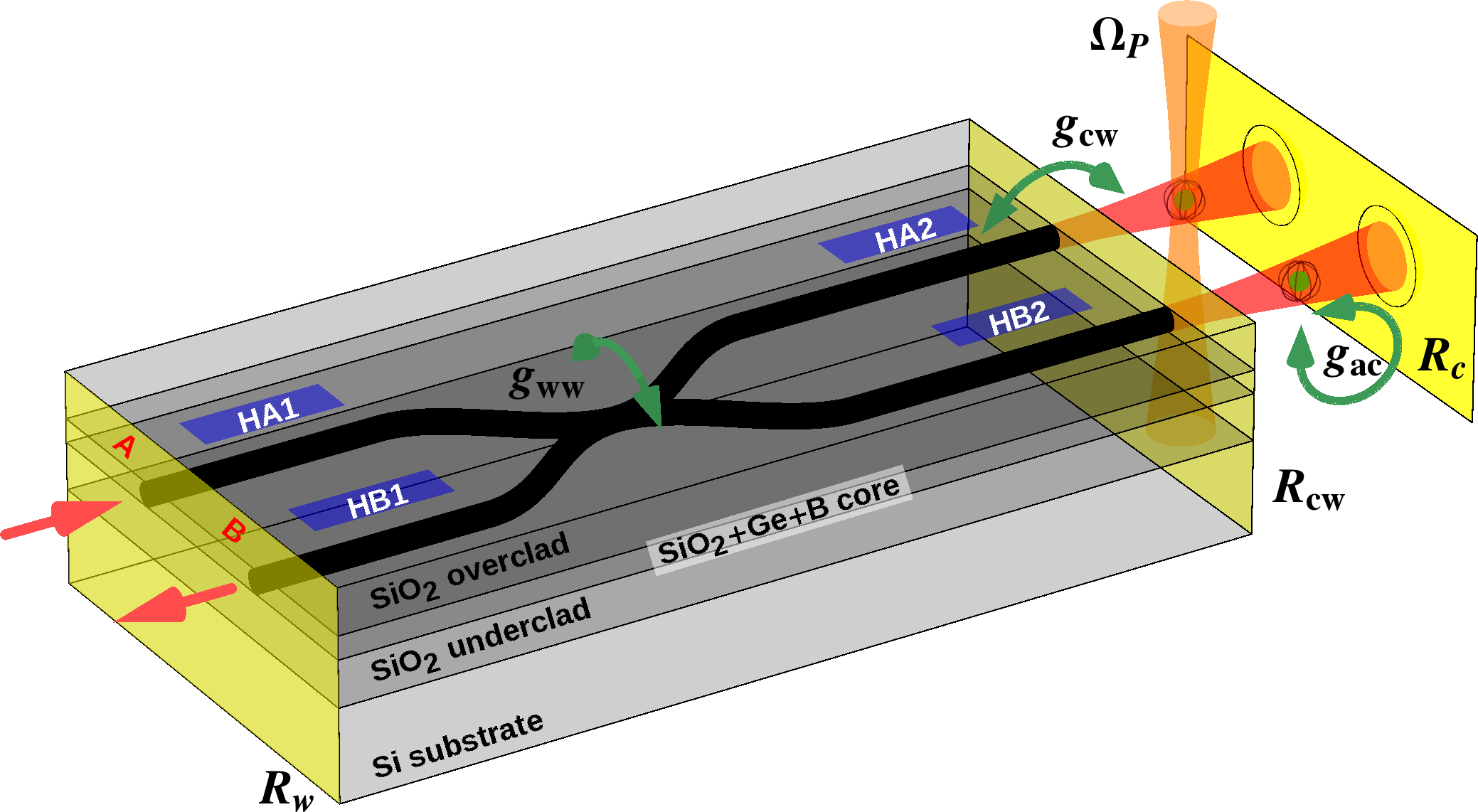}
	\caption{Sketch of the coupled-cavity QED structure discussed in \cite{Lepert2011njp} . Here we focus on the waveguide-resonator chip. $g\ac$, $g\cw$ and $g\ww$ are the three basic coupling rates, respectively atom-photon (i.e. vacuum Rabi frequency), microcavity-waveguide and waveguide-waveguide. Atoms can be manipulated with an external laser $\Omega_P$; waveguide lengths and $g\ww$ can be tuned with thermo-optic phase shifters (HA1 etc, in blue).}
	\label{fig:chip}
\end{figure}

This architecture, initially suggested in \cite{Poon2007}, appeared independently in our work on coupled-cavity quantum electrodynamics \cite{Lepert2011njp}. There we considered a set of evanescently coupled cavities, each formed by a \unit[\mytilde 1]{cm}-long waveguide terminated by mirrors of reflectivity $R\w$ and $R\cw$ coated on the end facets of the chip, as illustrated in Figure~\ref{fig:chip}. The purpose of these was to act as a bus, inter-connecting an array of free-space microcavities, each formed between one end of a waveguide and a spherical micro-mirror of reflectivity $R\c$ etched on silicon. The cavity-enhanced light-matter interaction (Purcell effect) in such a network has huge potential for quantum information processing \cite{Lepert2011njp}. All the major degrees of freedom are controllable: the atoms can be addressed with external lasers, the effective photon-photon interaction strength can be adjusted by detuning the microcavities, and  the waveguide-waveguide coupling rate $g\ww$ is tunable with the help of phase shifters (marked HA1 etc. in Fig.~\ref{fig:chip}). This tuneable coupling is the focus of our Letter.

Our waveguides, operating at at \unit[780]{nm}, were fabricated by UV writing as described in \cite{Lepert2011oe}. In the following experiments we used two types of coupler: evanescent and crossed. In the evanescent coupler, two waveguides come close together ($\unit[\mytilde 6]{\um}$) so that their modes overlap and can exchange energy. A 50:50 evanescent splitter requires a $\unit[\mytilde 1]{mm}$ length of coupling. In the crossed-coupler one waveguide is written on top of the other at a shallow angle, chosen to produce the desired coupling; $2.4^{\circ}$ yields a 50:50 splitting \cite{Kundys2009}. These are more compact than the evanescent couplers and display less variation in the coupling ratio, but are slightly more lossy (\unit[\mytilde 0.1]{dB}).

The mirrors are dielectric multilayers evaporated directly on the polished end facets of the chip \cite{OIB}. With a surface roughness of \unit[\mytilde 1]{nm}, reflectivities above 99.9\% are readily achievable. However, the propagation loss for our waveguides is $\unit[0.9\pm0.3]{dB/cm}$. In order to optimise the fringe visibility without compromising the finesse, we therefore choose mirror reflectivities of $R\w=92\%$ and $R\cw=99\%$. These give a finesse of $14\pm2$ for waveguides containing evanescent couplers, $12\pm2$ with crossed couplers, and $17\pm2$ for straight waveguides.

Phase shifters are deposited by sputtering $\unit[400]{nm}$-thick NiChrome wires on top of the waveguides. These are $ \unit[50]{\um}$ wide and  $\unit[1]{mm}$ long, with a typical resistance of $\unit[75]{\Omega}$.  Sputtered gold wires of similar dimensions link the NiCr wires to mm-wide gold pads which are connected to a power supply via spring-loaded gold pins mounted on a printed circuit board. The wires provide phase shifts of 3 to $5\pi$ per watt of dissipated power, with a typical response time of  \unit[0.5]{ms}. Normal operating conditions require less than \unit[0.5]{W}


\begin{figure}[tbp]
	\centering
	\includegraphics[width=\columnwidth]{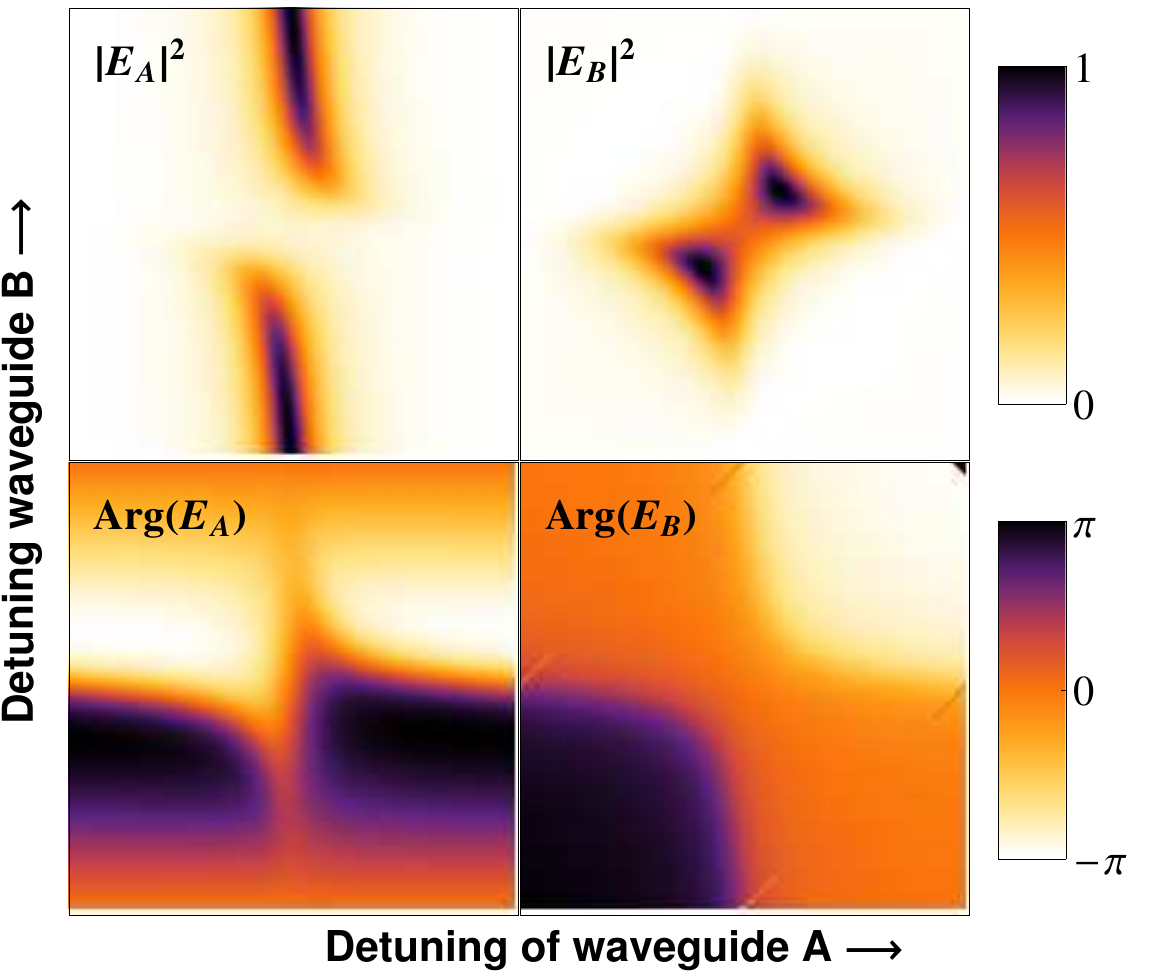}
	\caption{Theoretical intensity and phase of the fields circulating in two coupled waveguides resonators as a function of their lengths (detunings). The driving light is coupled into waveguide A. The plots cover $\pm \lambda/4$ on both axis, i.e. a full period and illustrate the symmetric/antisymmetric nature of the normal modes. The avoided crossing is clearly resolved.}
	\label{fig:WaveguideCav_Field+Phase}
\end{figure}

Consider an array of $N$ identical waveguides with equal couplings $\kappa$ \cite{Yariv}. Without mirrors, these support $N$ normal modes, superpositions of the individual waveguide modes that are invariant under propagation. Each of them, labelled $s=1,2,...N$, has a unique wavevector $\beta_s$ given by
\begin{equation}
	\beta_s=\frac{\omega n}{c} + 2\kappa\cos\frac{s\pi}{N+1}
	\label{eq:PropConst}
\end{equation}
where $c/n$ is the phase velocity of the guided mode of a single uncoupled waveguide. When waveguides of length $L$ are closed by mirrors, the $m^{th}$ resonance of mode  $s$ satisfies the resonance condition $\beta_s l + (\omega n/c) (L-l) = m \pi$, where $l$ is the length over which the waveguides are coupled. The cavity eigenfrequency that satisfies these two conditions simultaneously for transverse mode $s$ and longitudinal mode $m$ is 
\begin{equation}
	\omega_{m,s} = \frac{c/n}{L} \left(m \pi - 2 \kappa l\cos\frac{s \pi}{N+1}\right).
	\label{eq:crow_res}
\end{equation}
The first term is the resonant frequency of the uncoupled cavity. In the present case of $N=2$, the coupling shifts this frequency by $\pm \frac{c/n}{L} \kappa l$, corresponding to a coupling rate
\begin{equation}
	g\ww = \frac{c/n}{L} \kappa l.
	\label{eq:ww_coupling}
\end{equation}
Effectively, space-like or travelling-wave coupling has been converted to a time-like or standing-wave coupling by the addition of the resonators. It is important that only the product $\kappa \times l$ appears in Eq.~(\ref{eq:ww_coupling}); for $N=2$ it is related to the travelling-wave coupling ratio $C_{\times} = \sin^2( \kappa \times l)$, which can be easily measured before depositing the mirrors without any knowledge of the coupler parameters. The ``theoretical'' $g\ww$ of Eq.~(\ref{eq:ww_coupling}) can thus be compared to the experimental spectral splitting.

\begin{figure}[tbp]
	\centering
	\includegraphics[width=\columnwidth]{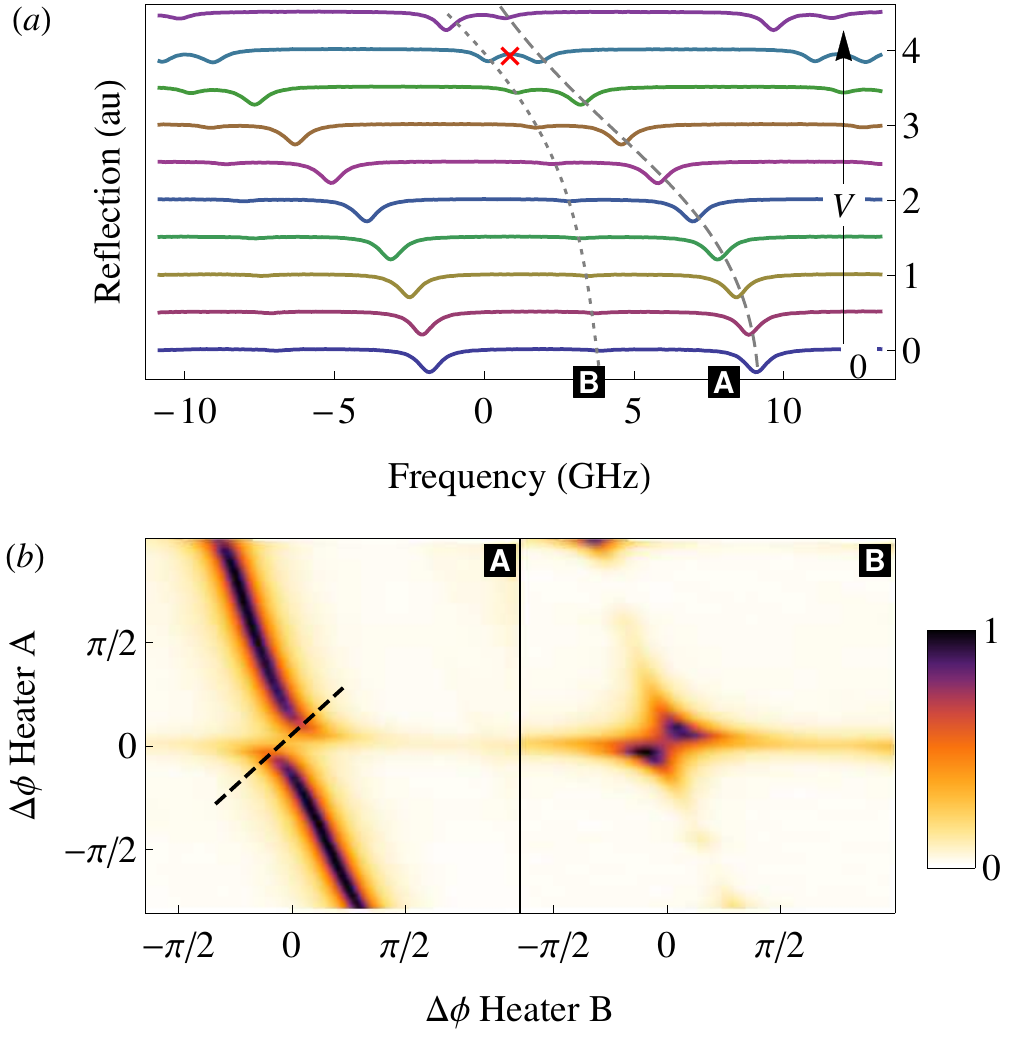}
	\caption{\emph{(a)} Evolution of the reflection spectrum from input arm A as its length increases. The spectra have been offset for clarity; $V$ is the voltage applied to heater HA1. Dashed line indicates resonance of waveguide A and dotted line that of B. \emph{(b)} Observation of avoided crossing around the red cross in top figure. Left: reflection at waveguide A. Right: output from waveguide B. Compare with the upper theoretical plots of Fig.~\ref{fig:WaveguideCav_Field+Phase}.}
	\label{fig:ORC8}
\end{figure}

\begin{figure}[btp]
	\centering
	\includegraphics[width=\columnwidth]{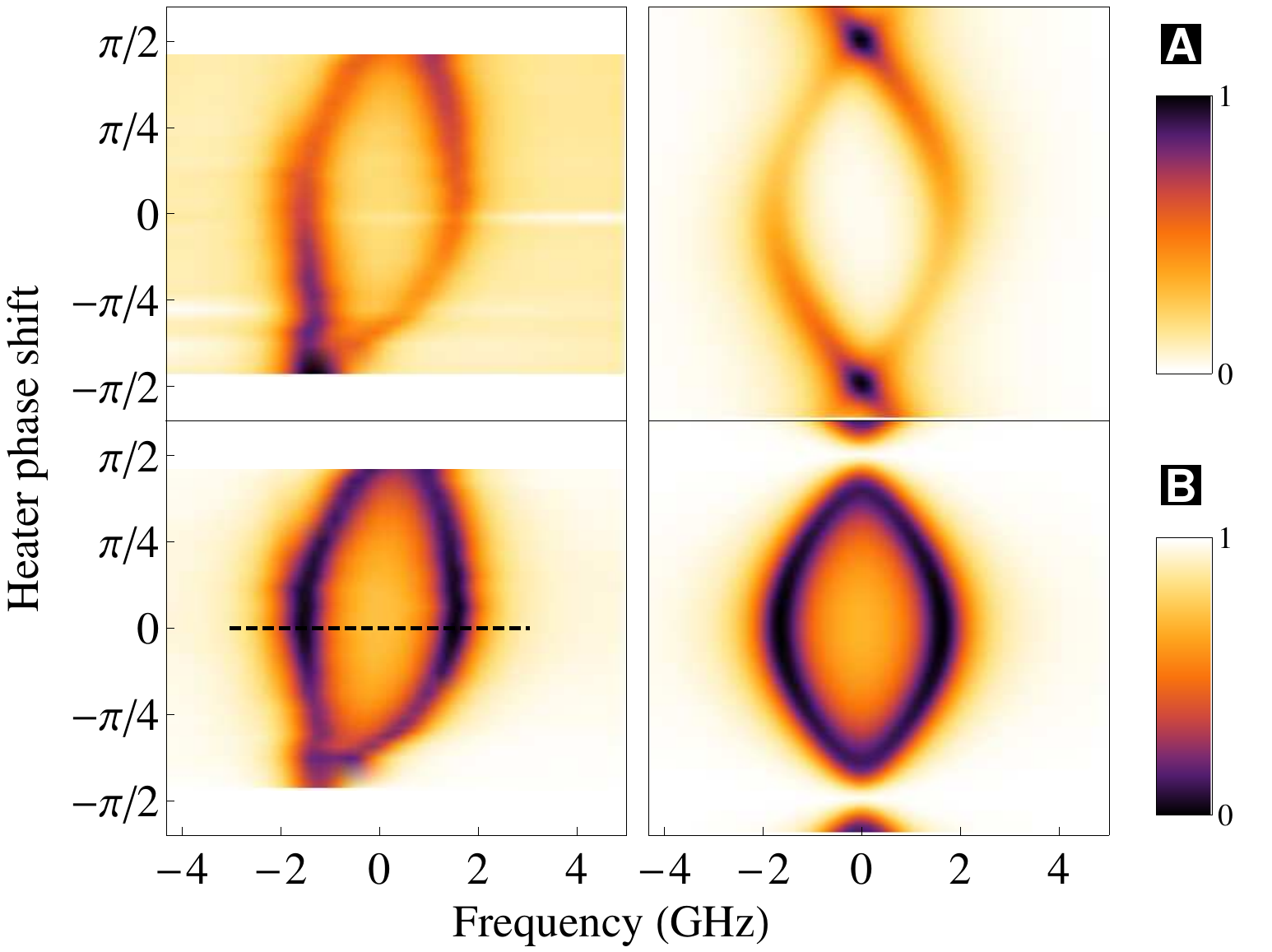}
	\caption{Tuning of the coupling rate $g\ww$. Top: experiment; bottom: theory. The plots show the output spectra from waveguides A and B as a function of the phase shifts induced by heaters HB1 and HB2.}
	\label{fig:ORC8-tuning}
\end{figure}

Standard transfer matrix methods can be used to calculate the whole spectrum \cite{Poon2007}. For the theoretical plots in Figs.~\ref{fig:WaveguideCav_Field+Phase} to \ref{fig:ORC8-tuning} we wrote the fields as an $N$-component vector, each component corresponding to the complex electric field amplitude of a different waveguide in the uncoupled region, and of a normal mode within the coupler. Together with basis changes and mirror reflections, this makes it possible to write a round trip matrix and to proceed with a summation of partially reflected wave, exactly as one would for a single FP resonator.

In Fig.~\ref{fig:WaveguideCav_Field+Phase} we map the magnitude and phase of the intracavity field as a function of both waveguide lengths (detunings). The upper and lower branches correspond to the two peaks of the split resonance. These branches are identical, except for the microcavity phases, which are opposite in sign. Thus we understand the two resonances as the symmetric and antisymmetric normal modes of two coupled oscillators. When the two normal modes are initially excited with the same amplitude, by placing a photon in one waveguide, the time evolution of the system will be a periodic exchange of the excitation between the two cavities, with a frequency $g\ww$ corresponding to half the normal mode splitting.

The coupling rate $g\ww$ in Eq.~(\ref{eq:ww_coupling}) can be tuned. It is proportional $\kappa$, which in turn is proportional to the overlap integral between the two coupled modes \cite{Yariv}. This depends on their relative phase $\psi$ as $\cos \psi$. Thus the tuning of $g\ww$ is as simple as lengthening one arm of the coupler and shortening the opposite arm to keep $L$ constant, which shifts $\psi$ while preserving the resonance condition. This is easily done using thermo-optic phase shifters.


Figure~\ref{fig:ORC8}(a) shows the experimental reflection spectrum of an evanescent coupler with a measured travelling-wave coupling ratio $C_{\times} = 18\pm2\%$. Laser light is injected into waveguide A with a butt-coupled optical fibre, and the reflected signal is collected through the same fibre. On the lowest trace, a single set of fringes is visible, corresponding to resonances of waveguide A. The length of the second waveguide B is slightly different from A due to fabrication imperfection, and is therefore not co-resonant: its resonances are barely perceptible on this trace. To bring A and B into resonance, we heat up heater HA1 (see Fig.~\ref{fig:chip}), thus increasing the length of waveguide A, so that resonance frequency A (dashed line) shifts to lower frequency. As this crosses resonance frequency B (dotted line; also shifting because of thermal cross-talk) it becomes strongly mixed. At $V \simeq \unit[4]{V}$, the two peaks have the same amplitude; A and B have the same length.

To show explicitly that this splitting corresponds to an avoided crossing, we adjust the laser frequency and heater voltages to put the system in the state marked by the red cross in Fig.~\ref{fig:ORC8}(a). The two waveguides are then co-resonant. We then scan the temperatures of HA1 and HB1 in order to change the lengths of both waveguides around this setpoint while the laser frequency remains fixed. The heaters have been individually calibrated so that the relationship between voltage and phase shift is known. The measured reflection from waveguide A and the measured output from B are plotted in Fig.~\ref{fig:ORC8}(b) as a function of the two heater phase shifts. The anti-crossing is well-resolved and closely resembles the theoretical plots in Fig.~\ref{fig:WaveguideCav_Field+Phase}. The tilt of the vertical resonance is caused, again, by thermal cross-talk as the slow scan of heater B changes the length of cavity A. The splitting measured along the dashed line in Fig.~\ref{fig:ORC8}(b) (left) is $g\ww = \unit[0.65 \pm 0.05]{GHz}$. With $C_{\times} = 18\pm2\%$, Eq.~\ref{eq:ww_coupling} gives $g\ww = \unit[1.4 \pm 0.1]{GHz}$, but there is no reason to expect the maximum coupling, since we have yet to tune the relative phase of the standing waves. 

	This we do now on a different device - an evanescent coupler having $C_{\times} = 30\pm2\%$. Starting again from a co-resonant configuration, and pumping on waveguide A, we translate the standing wave in waveguide B by cooling down HB1 while heating up HB2, in a balanced way to keep the overall waveguide length constant. The two waveguides remain co-resonant throughout. Experimental data are shown on the left of Fig.~\ref{fig:ORC8-tuning}, with simulations on the right. Initially there is almost no coupling (bottom of the graphs). The spectral splitting then increases to reaches a maximum of $g\ww = \unit[3.8 \pm 0.1]{GHz}$, measured along the dashed line, before decreasing again. The agreement with the theory, which gives $g\ww^{\mathrm{max}} = \unit[3.9 \pm 0.2]{GHz}$, is very good.

In conclusion, we have demonstrated a basic coupled-resonator optical waveguide in a side-coupled Fabry-Pérot configuration, in which the coupling rate can be conveniently tuned over a large range. The device is linearly scalable and can be straightforwardly interfaced with optical fibres or integrated into more complex waveguide chips, with potential applications as an optical delay line or as a key element in a cavity-QED based quantum simulator.

The authors acknowledge funding from the Royal Society, the EU (FP7: AQute, HIP) and the UK EPSRC.


\end{document}